\documentclass[english,iop]{emulateapj}
\usepackage[T1]{fontenc}
\usepackage[latin9]{inputenc}
\usepackage{color}
\usepackage{babel}

\usepackage{array}
\usepackage{booktabs}
\usepackage{units}
\usepackage{amsmath}
\usepackage{graphicx}
\usepackage{amssymb}

\makeatletter

\providecommand{\tabularnewline}{\\}

\makeatother

\begin{document}

\title{Terapixel imaging of cosmological simulations }

\author{Yu Feng\altaffilmark{1,2}, 
Rupert A.C. Croft\altaffilmark{1,2}, 
Tiziana Di Matteo\altaffilmark{1,2}, 
Nishikanta Khandai\altaffilmark{1,2}, 
Randy Sargent\altaffilmark{3}, 
Illah Nourbakhsh\altaffilmark{3}, 
Paul Dille\altaffilmark{3}, 
Chris Bartley\altaffilmark{3},
Volker Springel\altaffilmark{4,5},
Anirban Jana\altaffilmark{6}, Jeffrey Gardner\altaffilmark{7}}
\email{Email: yfeng1@andrew.cmu.edu}
\altaffiltext{1}{Bruce and Astrid McWilliams Center for Cosmology, Carnegie Mellon University, Pittsburgh, PA 15213}
\altaffiltext{2}{Department of Physics, Carnegie Mellon University, Pittsburgh, PA 15213}
\altaffiltext{3}{Robotics Institute, Carnegie Mellon University, Pittsburgh, PA 15213}
\altaffiltext{4}{Heidelberger Institut f\"{u}r Theoretische Studien, Schloss-Wolfsbrunnenweg 35, 69118 Heidelberg, Germany}
\altaffiltext{5}{Zentrum f\"ur Astronomie der Universit\"at Heidelberg, Astronomisches Recheninstitut, M\"{o}nchhofstr. 12-14, 69120 Heidelberg, Germany}

\altaffiltext{6}{Pittsburgh Supercomputing Center, Pittsburgh, PA 15213 }
\altaffiltext{7}{Physics Department, University of Washington, Seattle, WA 98195}

\keywords{Cosmology: observations -- large-scale structure of Universe}
\begin{abstract}
The increasing size of cosmological simulations has led to
the need for new visualization techniques.  We focus on
Smoothed Particle Hydrodynamical (SPH) simulations run with
the {\small GADGET} code and describe methods for visually
accessing the entire simulation at full resolution.  The
simulation snapshots are rastered and processed on
supercomputers into images that are ready to be accessed
through a  web interface (GigaPan).  This allows any
scientist with a web-browser to interactively explore
simulation datasets in both in spatial and temporal
dimensions, datasets which in their native format can be
hundreds of terabytes in size or more. We present two
examples, the first a static terapixel image of the
MassiveBlack simulation, a {\small P-GADGET} SPH simulation
with 65 billion particles, and the second an interactively
zoomable animation of a different simulation with more than
one thousand frames, each a gigapixel in size. Both are
available for public access through the GigaPan web
interface.  We also make our imaging software publicly
available.
\end{abstract}
\maketitle

\section{Introduction}

In the 40 years that $N$-body simulations have been used in
Cosmology research, visualization has been the most
indispensable tool.  Physical processes have often been
identified first and studied  via images of simulations. A
few examples are: formation of filamentary structures in the
large-scale distribution of matter
 \citep{JENKINS,COPLBERG,MILLENIUM-I,DOLAG}, growth of
feedback bubbles around quasars
 \citep{SIJACKI,FEEDBACK}; cold flows of gas forming
galaxies  \citep{COLDFLOW-DEKELA,KERES}, and the
evolution of ionization fronts during the re-ionization
epoch \citep{BUBBLESHIN,BUBBLEZAHN}. The size of current and
upcoming Peta-scale simulation datasets can make such visual
exploration to discover new physics technically challenging.
Here we present techniques that can  be used to display
images at full resolution of datasets of hundreds of
billions of particles in size.

Several implementations of visualization software for
cosmological simulations already exist.
IRFIT \citep{IFRIT} is a general purpose visualization
suite that can deal with mesh based scalar, vector and
tensor data, as well as particle based datasets as points.
YT \citep{YT} is an analysis toolkit for mesh based
simulations that also supports imaging.
SPLASH \citep{SPLASH} is a visualization suite
specialized for simulations that use smoothed particle
hydrodynamics(SPH) techniques. Aside from the CPU based
approaches mentioned above, \cite{GPUVIS} implemented a
GPU based interactive visualization tool for SPH
simulations.

The Millennium I \& II simulations
 \citep{MILLENIUM-I,MILLENIUM-II} have been used to test
an interactive scalable rendering system developed by
\citet{FSW2009}. Both SPLASH and the Millennium visualizer
support high quality visualization of SPH data sets, while
IRFIT treats SPH data as discrete points.

Continuing improvements in  computing technology and
algorithms are allowing  SPH cosmological simulations to be
run with ever increasing numbers of particles.  Runs are now
possible on scales which allow rare objects, such as quasars
to form in a large simulation volume with uniform high
resolution (see Section 2.1; and
\citet{DIMATTEO-PREP,DEGRAF-PREP,KHANDAI-PREP}).  Being
able to scan through a vast volume and seek out the tiny
regions of space where most of the activity is occurring,
while still keeping the large-scale structure in context
necessitates special visualization capabilities. These
should be able to show the largest scale information but at
the same be interactively zoomable.  However, as the size of
the datasets quickly exceeds the capability of moderately
sized in-house computer clusters, it becomes difficult to
perform any interactive visualizations. For example, a
single snapshot of the MassiveBlack simulation (Section 2.1)
consists of 8192 files and is over 3 TB in size. 

Even when a required large scale high resolution image has
been rendered, actually exploring the data requires special
tools.  The GigaPan collaboration
\footnote{http://www.gigapan.org} has essentially solved
this problem in the context of viewing large images, with
the GigaPan viewer enabling anyone connected to the Internet
to zoom into and explore in real time images which would
take hours to transfer in totality. The viewing technology
has been primarily used to access large photographic
panoramas, but is easily applicable to simulated datasets. A
recent enhancement to deal with the time dimension, in the
form of gigapixel frame interactive movies (GigaPan Time
Machine\footnote{http://timemachine.gigapan.org}) will turn
out to give particularly novel and exciting results when
applied to simulation visualization.

In this work we combine an off-line imaging technique
together with GigaPan technology to implement an
interactively accessible  visual probe of large cosmological
simulations. While GigaPan is an independent project
(uploading and access to the GigaPan website is publicly
available), we release our  toolkit for the off-line
visualization  as {\small Gaepsi}%
\footnote{http://web.phys.cmu.edu/\textasciitilde{}yfeng1/gaepsi%
}, a software package aimed specifically at {\small GADGET}
 \citep{GADGET,GADGET2} SPH simulations. 

The layout of our paper is as follows. In Section 2 we give
a brief overview of the physical processes modeled in
{\small GADGET}, as well as describing two {\small P-GADGET}
simulations which we have visualized. In Section 3 we give
details of the spatial domain remapping we employ to convert
cubical simulation volumes into image slices. In Section 4,
we describe the process of rasterizing an SPH density field,
and in Section 5  the image rendering and layer compositing.
In Section 6 we address the parallelism of our techniques
and give measures of performance. In Section 7 we briefly
describe the GigaPan and GigaPan Time Machine viewers and
present examples screenshots from two visualizations (which
are both accessible on the GigaPan websites).

\section{Simulation}

Adaptive Mesh Refinement(AMR, e.g., \citet{AMR}) and
Smoothed Particle Hydrodynamics (SPH, \citet{SPH}) are the
two most used schemes for carrying out cosmological
simulations. In this work we focus on the visualization of
the baryonic matter in SPH simulations run with {\small
P-GADGET}  \citep{GADGET2}.

 {\small GADGET} is an SPH implementation, and {\small
P-GADGET} is a version which has been developed specifically
for petascale computing resources. It simultaneously
follows the self-gravitation of a collision-less N-body
system (dark matter) and gas dynamics (baryonic matter), as
well as the formation of stars and super-massive black holes.
Dark matter particles and gas particle positions and
initial characteristics are set up in a comoving cube, and
black hole and star particles are created according to
sub-grid modeling  \citep{STARFORMATION, BLACKHOLE} Gas
particles carry  hydrodynamical properties, such as
temperature, star formation rate, and neutral fraction.

Although our attention in this paper is limited to imaging
properties the of gas, stars and black holes in {\small
GADGET} simulations, similar techniques could be used to
visualize the dark matter content.  Also, the software we
provide should be easily adaptable to the data formats of
other SPH codes (e.g. {\small GASOLINE}, \citep{GASOLINE})

\subsection{MassiveBlack}

The MassiveBlack simulation is the state-of-art SPH
simulation of a  $\mathsf{\Lambda CDM}$
universe \citep{DIMATTEO-PREP}. {\small P-GADGET} was
used to evolve  $2\times3200^{3}$ particles in a volume of
side length $\unit[533]{h^{-1}Mpc}$ with a gravitational
force resolution of $\unit[5]{h^{-1}Kpc}$.  One snapshot of
the simulation occupies 3 tera-bytes of disk space, and the
simulation has been run so far to redshift $z=4.75$,
creating a dataset of order $120$ TB.  The fine resolution
and large volume of the simulation permits one to usefully
create extremely large images.  The simulation was run on
the high performance computing facility, Kraken, at the
National Institute for Computational Sciences in full
capability mode with 98,000 CPUs.%

\subsection{E5}

To make a smooth animation of the evolution of the universe
typically requires hundreds frames directly taken as
snapshots of the simulation. The scale of the MassiveBlack
run is too large for this purpose, so we ran a much smaller
simulation (E5) with $2\times336^{3}$ particles in a
$\unit[50]{h^{-1}Mpc}$ comoving box.  The model was again a
$\mathsf{\Lambda CDM}$ cosmology, and one snapshot  was
output per 10 million years, resulting in 1367 snapshots.
This simulation ran on $256$ cores of the Warp cluster in
the McWilliams Center for Cosmology at CMU.

\section{Spatial Domain Remapping}

Spatial domain remapping can be used to transform the
periodic cubic domain of a cosmological simulation to a
patch whose shape is similar to the domain of a sky survey,
while making sure that the local structures in the
simulation are preserved  \citep{REMAPPING,REMAPPING2}.
Another application is making a thin slice that includes the
entire volume of the simulation.  Our example will focus on
the latter case. 

\begin{figure}
\centering
\includegraphics[scale=0.9]{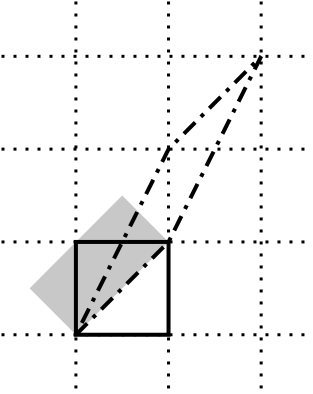}

\caption{Transformation of the Primitive
Cell\label{fig:TransformationPrimitive}.  The cubical unit
cell is shown using solid lines. The new primitive cell,
generated by $\left[\protect\begin{smallmatrix} 1 &
1\protect\\ 1 & 2\protect\end{smallmatrix}\right]$ is shown with
dash-dotted lines.  The transformed domain is shown in gray.
}

\end{figure}

A {\small GADGET} cosmological simulation is usually defined
in the periodic domain of a cube. As a result, if we let
$f(X=(x,y,z))$ be any position dependent property of the
simulation, then \[ f(X)=f((x+\mu L,y+\nu L,z+\sigma L)),\]
where $\mu,\nu,\sigma$ are integers. The structure
corresponds to a simple cubic lattice with lattice constant
$a=L$, the simulation box side-length. A bijective mapping
from the cubic unit cell to a remapped domain corresponds to
a choice of the primitive cell. Figure
\ref{fig:TransformationPrimitive} illustrates the situation
in 2 dimensions. %

Whilst the original remapping algorithm by
\cite{REMAPPING} results in the correct transformations
being applied, it has two drawbacks: (i) the
orthogonalization is invoked explicitly and (ii) the
hit-testing for calculation of the shifting (see below) is
against non-aligned cuboids. The second problem especially
undermines the performance of the program. In this
work we present a faster algorithm based on similar ideas,
but which features a QR decomposition (which is widely
available as a library routine), and hit-testing against an
AABB (Axis Aligned Bounding Box).

First, the transformation of the primitive cell is given by
a uni-modular integer matrix, \[
M=\left(\begin{matrix}
M_{11} & M_{12} & M_{13}\\
M_{21} & M_{22} & M_{23}\\
M_{31} & M_{32} & M_{33}\end{matrix}\right),\]
where $M_{ij}$ are integers and the determinant of the
matrix $|M|=1$.  It is straight-forward to obtain such
matrices via enumeration. \citep{REMAPPING} The $QR$
decomposition of $M$ is \[ M=QR,\] where $Q$ is an
orthonormal matrix and $R$ is an upper-right triangular
matrix. It is immediately apparent that (i) application of
$Q$ yields rotation of the basis from the simulation domain
to the transformed domain, the column vectors in matrix
$Q^{T}$ being the lattice vectors in the transformed domain;
(ii) the diagonal elements of $R$ are the dimensions of the
remapped domain. For imaging it is desired that the
thickness along the line of sight is significantly shorter
than the extension in the other dimensions, thus we require
$0<R_{33}\ll|R_{22}|<|R_{11}|$.  Note that if a domain that
is much longer in the line of sight direction is desired,
for example to calculate long range correlations or to make
a sky map of a whole simulation in projection, the choice
should be $0<|R_{33}|<|R_{22}|\ll|R_{11}|$. 

Next, for each sample position $X$, we solve the indefinite
equation of integer cell number triads
$I=(I_{1},I_{2},I_{3})^{T}$,\begin{flalign*} \tilde{X} &
=Q^{T}X+aQ^{T}I,\end{flalign*} where $a$ is the box size,
$\tilde{X}$ is the transformed sample position satisfying
$\tilde{X}\in[0,R_{11})\times[0,R_{22})\times[0,R_{33})$.
In practice, the domain of $\tilde{X}$ is enlarged by a
small number $\epsilon$ to address numerical errors.
Multiplying by $Q$ on the left and re-organizing the terms,
we find \[ I=\frac{Q\tilde{X}}{a}-\frac{X}{a}.\] Notice that
$Q\tilde{X}$ is the transformed sample position expressed in
the original coordinate system, and is bounded by its AABB
box.  If we let $(Q\tilde{X}/a)_{i}\in[B_{i},T_{i}]$, where
$B_{i}$ and $T_{i}$ are integers, and notice
$\frac{X_{i}}{a}\in[0,1)$, the resulting bounds of $I$ are
given by \[ I_{i}\in[B_{i},T_{i}].\] We then enumerate the
range to find $\tilde{X}$.

\begin{figure}
\centering
\includegraphics[scale=0.7]{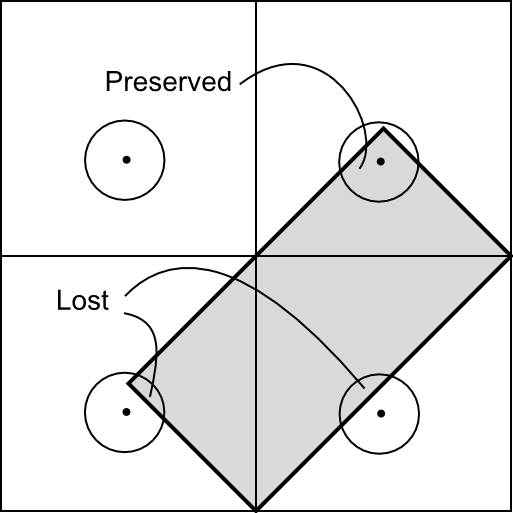}\caption{Boundary
effects and Smoothed Particles\label{fig:Distortion}. Four
images of a particle intersecting the boundary are shown.
The top-right image is contained in the transformed domain,
but the other three are not. The contribution of the two
bottom images is lost.  By requiring the size of the
transformed domain to be much larger than typical SPH
smoothing lengths, most particles do not intersect a
boundary of the domain and the error is contained near the
edges.}

\end{figure}

When the remapping method is applied to the SPH particle
positions, the transformations of the particles that are
close to the edges give inexact results. The situation is
similar to the boundary error in the original domain when
the periodic boundary condition is not properly considered.
Figure \ref{fig:Distortion} illustrates the situation by
showing all images of the particles that contribute to the
imaging domain. We note that for the purpose of imaging, by
choosing a $R_{33}$(the thickness in the thinner dimension)
much larger than the typical smoothing length of the SPH
particle, the errors are largely constrained to lie near the
edge. These issues are part of general complications related
to the use of a simulation slice for visualization. For
example in an animation of the distribution of matter in a
slice it is possible for objects to appear and disappear in
the middle of the slice as they pass through it. These
limitations should be borne in mind, and we leave 3D visualization
techniques for future work.

The transformations used for the MassiveBlack and E5
simulations are listed in Table \ref{tab:Transformations}.
\begin{table}
\caption{Transformations\label{tab:Transformations}}
\small
\centering\begin{tabular}{ccc}
\toprule 
Simulation & MassiveBlack & E5\tabularnewline
\midrule
\midrule 
Matrix $M$ & $\left[\begin{smallmatrix}
5 & 6 & 2\\
3 & -7 & 3\\
2 & 7 & 0\end{smallmatrix}\right]$ &
$\left[\tiny\begin{smallmatrix}
3 & 1 & 0\\
2 & 1 & 0\\
0 & 2 & 1\end{smallmatrix}\right]$\tabularnewline
\midrule 
$R_{11}$(h$^{-1}$Mpc) & 3300 & 180\tabularnewline
\midrule 
$R_{11}$ Pixels(Kilo) & 810 & 36.57\tabularnewline
\midrule 
$R_{22}$(h$^{-1}$Mpc) & 5800 & 101\tabularnewline
\midrule 
$R_{22}$ Pixels(Kilo) & 1440 & 20.48\tabularnewline
\midrule 
$R_{33}$(h$^{-1}$Mpc) & 7.9 & 6.7\tabularnewline
\midrule 
Resolution(h$^{-1}$Kpc) & 4.2 & 4.9\tabularnewline
\bottomrule
\end{tabular}
\end{table}

\section{Rasterization}

In a simulation, many field variables are of interest in
visualization.  \begin{itemize} \item scalar fields :
density $\rho$, temperature $T$, neutral fraction
$x_{\text{HI}}$, star formation rate $\phi$; \item vector
fields: velocity, gravitational force.  \end{itemize} In an
SPH simulation, a field variable as a function of spatial
position is given by the interpolation of the particle
properties. Rasterization converts the interpolated
continuous field into raster pixels on a uniform grid. The
kernel function of  a particle at position $\mathbf{y}$ with
smoothing length $h$ is defined as \[
W(\mathbf{y},h)=\frac{8}{\pi h^{3}}\begin{cases}
1-6\left(\frac{y}{h}\right)^{2}+6 &
\left(\frac{y}{h}\right)^{3},0\le\frac{y}{h}\le\frac{1}{2}\\
2\left(1-\frac{y}{h}\right)^{3}, &
\frac{1}{2}<\frac{y}{h}\le1,\\ 0 &
\frac{y}{h}>1.\end{cases}\] Following the usual
prescriptions \citep[e.g.][]{GADGET2,SPLASH}, the interpolated
density field is taken as\[
\rho(\mathbf{x})=\sum_{i}m_{i}W(\mathbf{x}-\mathbf{x}_{i},h_{i}),\]
where $m_{i}$, $\mathbf{x}_{i}$, $h_{i}$ are the mass,
position, and smoothing length of the $i$th particle,
respectively. The interpolation of a field variable, denoted
by $A$, is given by \[
A(\mathbf{x})=\sum_{i}\frac{A_{i}m_{i}W(\mathbf{x}-\mathbf{x}_{i},h_{i})}{\rho(\mathbf{x}_{i})}+O(h^{2}),\]
where $A_{i}$ is the corresponding field property carried by
the $i$th particle. Note that the density field can be seen
as a special case of the general formula.

Two types of pixel-wise mean for a field are calculated, the
\begin{enumerate} 
\item volume-weighted mean of the density field, 
\begin{flalign*} \bar{\rho}(P) &
=\frac{M(P)}{P}
=\frac{1}{P}\int_{P}d^{3}\mathbf{x}\,\rho(\mathbf{x})\\
& =\frac{1}{P}\sum_{i}m_{i}\int_{P}d^{3}\mathbf{x}\,
W(\mathbf{x}-\mathbf{x}_{i},h_{i})\\ &
=\frac{1}{P}\sum_{i}M_{i}(P),
\end{flalign*} 
where
$M_{i}(P)=m_{i}\int_{P}d^{3}\mathbf{x}\,
W(\mathbf{x}-\mathbf{x}_{i},h_{i})$ is the mass overlapping
of the $i$th particle and the pixel, and the \item
mass-weighted mean of a field ($A$)
\footnote{The second line is an approximation. For the
numerator,
\begin{flalign*} 
& \int_{P}d^{3}\mathbf{x}\, A(\mathbf{x})\rho(\mathbf{x})=\\ 
& \int_{P}d^{3}\mathbf{x}\,\sum_{i}\frac{A_{i}m_{i}W(\mathbf{x}-\mathbf{x}_{i},h_{i})}{\rho_{i}}\sum_{j}m_{j}W(\mathbf{x}-\mathbf{x}_{j},h_{j})\\
&
=\sum_{i}\frac{A_{i}m_{i}}{\rho_{i}}\sum_{j}m_{j}\int_{P}d^{3}\mathbf{x}\,
W(\mathbf{x}-\mathbf{x}_{i},h_{i})W(\mathbf{x}-\mathbf{x}_{j},h_{j}).\end{flalign*}
If we apply the mean value theorem to the integral and
Taylor expand, we find that \begin{flalign*} &
\int_{P}d^{3}\mathbf{x}\,
W(\mathbf{x}-\mathbf{x}_{i},h_{i})W(\mathbf{x}-\mathbf{x}_{j},h_{j})\\
&
=W(\mathbf{\xi}_{ij}-\mathbf{x}_{j},h_{j})\int_{P}d^{3}\mathbf{x}\,
W(\mathbf{x}-\mathbf{x}_{i},h_{i})\\ = &
\left[W(\mathbf{x}_{i}-\mathbf{x}_{j},h_{j})+W'(\mathbf{x}_{i}-\mathbf{x}_{j},h_{j})(\mathbf{\xi}_{ij}-\mathbf{x}_{i})\right.\\
& +\left.O[(\mathbf{\xi}_{ij}-\mathbf{x}_{i})^{2}]\right]\\
& \times\int_{P}d^{3}\mathbf{x}\,
W(\mathbf{x}-\mathbf{x}_{i},h_{i})\\ &
=W(\mathbf{x}_{i}-\mathbf{x}_{j},h_{j})\int_{P}d^{3}\mathbf{x}\,
W(\mathbf{x}-\mathbf{x}_{i},h_{i})+O(h^{2}).\end{flalign*}
Both $W'$ and $\mathbf{\xi}_{ij}-\mathbf{x}_{i}$ are bound
by terms of  $O(h_{j})$, so that the extra terms are all
beyond $O(h^{2})$ (the last line).  Noticing that
$\rho_{i}=\sum_{j}m_{j}W(\mathbf{x}_{i}-\mathbf{x}_{j},h_{j})+O(h^{2})$,
the numerator \begin{flalign*} & \int_{P}d^{3}\mathbf{x}\,
A(\mathbf{x})\rho(\mathbf{x})\\ &
=\sum_{i}A_{i}m_{i}\int_{P}d^{3}\mathbf{x}\,
W(\mathbf{x}-\mathbf{x}_{i},h_{i})+O(h^{2})\\ &
=\sum_{i}A_{i}M_{i}(P)+O(h^{2}).\end{flalign*}
},
\begin{eqnarray*}
\bar{A}(P) & = & \frac{\int_{P}d^{3}\mathbf{x}\,
A(\mathbf{x})\rho(\mathbf{x})}{\int_{P}d^{3}\mathbf{x}\,\rho(\mathbf{x})}\\
& = &
\frac{\sum_{i}A_{i}M_{i}(P)}{M(P)}+O(h^{2}).\end{eqnarray*}
\end{enumerate}

To obtain a line of sight projection along the third axis,
the pixels are chosen to extend along the third dimension,
resulting a two dimensional final raster image. The
calculation of the overlapping $M_{i}(P)$ in this
circumstance is two dimensional. Both formulas require
frequent calculation of the overlap between the kernel
function and the pixels. An effective way to calculate the
overlap is via a lookup table that is pre-calculated and
hard coded in the program. Three levels of approximation are
used in the calculation of the contribution of a particle to
a pixel:
\begin{enumerate} \item When a particle is much smaller than
a pixel, the particle contributes to the pixel as a whole.
No interpolation and lookup occurs.  \item When a particle
and pixel are  of similar size of (up to a few pixels in
size), the contribution to each of the pixels are calculated
by interpolating between the overlapping areas read from a
lookup table.  \item When a particle is much larger than a
pixel, the contribution to a pixel taken to be the center
kernel value times the area of a pixel.  \end{enumerate}
Note that Level 1 and 3 are significantly faster than Level
2 as they do not require interpolations.

The rasterization of the $z=4.75$ snapshot of MassiveBlack was
run on the  SGI UV Blacklight supercomputer at the Pittsburgh
Supercomputing Center.  Blacklight is a shared memory
machine equipped with a large memory for holding the image
and a fairly large number of cores enabling parallelism,
making it the most favorable machine for the rasterization.
The rasterization of the E5 simulation was run on local CMU
machine Warp. The pixel dimensions of the raster images are
also listed in Table \ref{tab:Transformations}. The pixel
scales have been  chosen to be around the gravitational
softening length of $\sim \unit[5]{h^{-1}Kpc}$ in these
simulations in order to preserve as much information in the
image as possible.

\section{Image Rendering and Layer Compositing}

The rasterized SPH images are color-mapped into RGBA (red,
green, blue and opacity) layers. Two modes of color-mapping
are implemented, the simple mode and the intensity mode. 

In the simple mode, the color of a pixel is directly
obtained by looking up the normalized pixel value in a given
color table. To address the large (several orders of
magnitude) variation of the fields, the logarithm of the
pixel value is used in place of the pixel value itself. 

In the intensity mode, the color of a pixel is determined in
the same way as done in the simple mode. However, the
opacity is reduced by a factor $f_{m}$ that is determined by
the logarithm of the total mass of the SPH fluid contained
within the pixel. To be more specific, \[
f_{m}=\begin{cases} 0 & \log M<a,\\ 1 & \log M>b,\\
\left[\frac{\log M-a}{b-a}\right]^{\gamma} &
\text{otherwise},\end{cases}\] where $a$ and $b$ are the
underexposure and overexposure parameters: any pixel that
has a mass below $10^{a}$ is completely transparent, and any
pixel that has a mass above $10^{b}$ is completely opaque. 

The RGBA layers are stacked one on top of another to
composite the final image. The compositing assumes an opaque
black background.  The formula to composite an opaque
bottom layer $B$ with an overlay layer $T$ into the
composite layer $C$ is \citep{PTDT1984} 
\[ C=\alpha
F+(1-\alpha)T,
\] 
where $C$, $B$ and $T$ stand for the RGB
pixel color triplets of the corresponding layer and $\alpha$
is the opacity value of the pixel in the overlay layer $T$.
For example, if the background is red and the overlay color
is green, with $\alpha=50\%$, the composite color is a
$50\%$-dimmed yellow. 

Point-like (non SPH) particles are rendered differently.
Star particles are rendered as colored points, while black
hole particles are rendered using circles, with the radius
proportional to the logarithm of the mass. In our example
images, the MassiveBlack simulation visualization used  a
fast rasterizer that does not support anti-aliasing, whilst
the frames of E5 are rendered using
matplotlib \citep{MATPLOTLIB} that does anti-aliasing.

\begin{figure}
\centering
\includegraphics[width=2in]{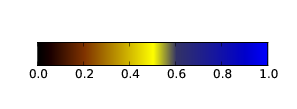}

\caption{Fiducial color-map for the
gas density field\label{fig:Colormap-of-gas}. The colors
span a darkened red through yellow to blue.}

\end{figure}

The choice of the colors in the color map has to be made
carefully to avoid confusing different quantities.  We
choose a color gradient which  spans  black, red, yellow and
blue for the color map of the normalized gas density field.
This color map is shown in Figure \ref{fig:Colormap-of-gas}.
Composited above the gas density field is the mass weighted
average of the star formation rate field, shown in dark
blue, and with completely transparency where the field
vanishes.  Additionally, we choose solid white pixels for
the star particles.  Blackholes are shown as green circles.
In the E5 animation frames, the normalization of the gas density
color map has been fixed so that the maximum and minimum
values correspond to the extreme values of density in the
last snapshot. %

\section{Parallelism and Performance}

The large simulations we are interested in visualizing have
been run on large supercomputer facilities. In order to
image them with sufficient resolution to be truly useful,
the creation of images from the raw simulation data also
needs significant computing resources. In this section we
outline our algorithms for doing this and give measures of
performance. 

\subsection{ Rasterization in parallel}

We have implemented two types of parallelism, which we shall
refer to as ``tiny'' and ``huge'', to make best use of
shared memory architectures and distributed memory
architectures, respectively. The tiny parallelism is
implemented with OpenMP and takes advantage of the case when
the image can be held within the memory of a single
computing node.  The parallelism is achieved by distributing
the particles in batches to the threads within one computing
node. The raster pixels are then color-mapped in serial, as
is the drawing of the point-like particles.  The tiny mode
is especially useful for interactively probing smaller
simulations.

The huge version of parallelism is implemented using the
Message Passing Interface (MPI) libraries and is used when
the image is larger than a single computing node or the
computing resources within one node are insufficient to
finish the rasterization in a timely manner. The imaging
domain is divided into horizontal stripes, each of which is
hosted by a computing node. When the snapshot is read into
memory, only the particles that contribute to the pixels in
a domain are scattered to the hosting node of the domain. Due
to the growth of cosmic structure as we move to lower
redshifts, some of the stripes inevitably have many more
particles than others, introducing load imbalance.  We
define the load imbalance penalty $\eta$ as the ratio
between the maximum and the average of the number of
particles in a stripe. The computing nodes with fewer
particles tend to finish sooner than those with more. The
color-mapping and the drawing of point-like particles are
also performed in parallel in the huge version of
parallelism.

\subsection{Performance}

The time spent in domain remapping scales linearly with the
total number of particles $N$,\[ T_{\text{remap}}\sim
O(N).\] The time spent in color-mapping scales linearly with
the total number of pixels $P$,\[ T_{\text{color}}\sim
O(P).\] Both processes consume a very small fraction of the
total computing cycles. 

The rasterization consumes a much  larger part of the
computing resources and it is useful to analyze it in more
detail. If we let $\bar{n}$ be the number of pixels
overlapping a particle, then $\bar{n}=K^{-1}N^{-1}P$, where
$K^{-1}$ is a constant related to the simulation.  Now we
let $t(n)$ be the time it takes to rasterize one particle,
as a function of the number of pixels overlapping the
particle. From the 3 levels of detail in the rasterization
algorithm (Section 4), we have \[ t(n)=\begin{cases} C_{1},
& n\ll1\\ C_{2}n & n\sim1\\ C_{3}n, & n\gg1\end{cases};\]
with $C_{2}\gg C_{1}\approx C_{3}$ . The effective pixel
filling rate $R$ is defined as the total number of image
pixels rasterized per unit time,\begin{gather*}
R=P[t(\bar{n)}N]^{-1}=\bar{n}K[t(\bar{n})]^{-1}\\
=\begin{cases} \bar{n}KC_{1}^{-1}, & \bar{n}\ll1\\
KC_{2}^{-1} & \bar{n}\sim1\\ KC_{3}^{-1}, &
\bar{n}\gg1\end{cases}.\end{gather*}

\begin{figure}
\centering
\includegraphics[width=2.7in]{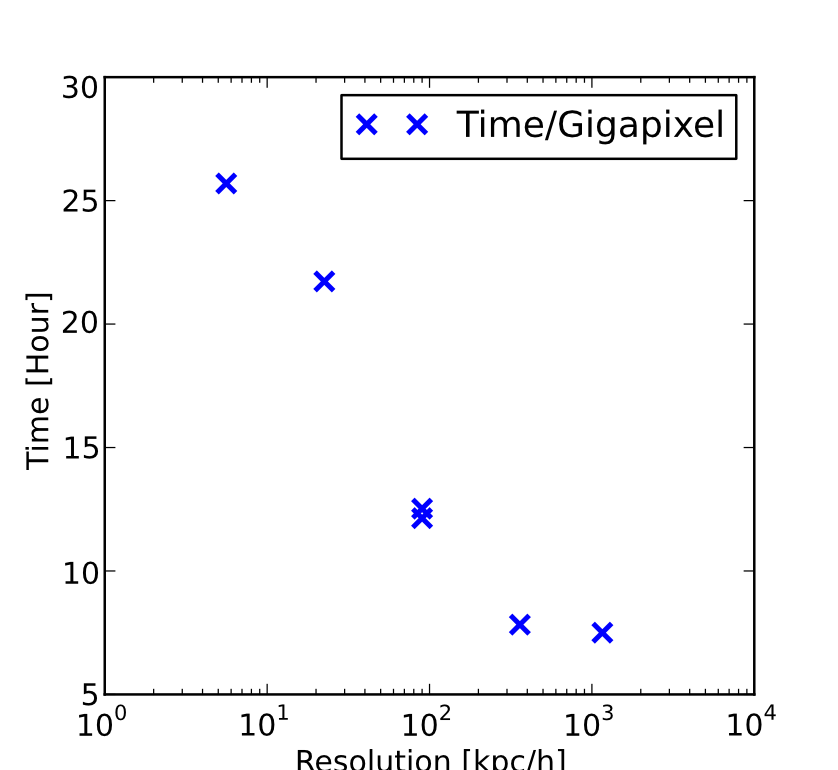}

\caption{MassiveBlack Simulation Rasterization
Rate\label{fig:RasterizationRate}. We show the number of
pixels fixed as a function of resolution (pixel scale). The
rate peaks at $KC_{3}^{-1}$ at the high resolution limit and
approaches $KC_{2}^{-1}$ as the resolution worsens. The
$\bar{n}\ll1$ domain is not explored.}

\end{figure}

The rasterization time to taken to create images from a
single snapshot of the MassiveBlack simulation (at redshift
4.75) at various resolutions is presented in Table
\ref{tab:RasterizationTime} and Figure
\ref{fig:RasterizationRate}. %

\begin{table}
\small
\caption{Time Taken to Rasterize the MassiveBlack Simulation
\label{tab:RasterizationTime}. Here $N$ is the number of SPH
particles, Res is the resolution (pixel scale),
$\overline{n}$ is mean number of pixels that overlap each
particle, $\eta$ is the load unbalance penalty averaged over
patches, and the final column, Rate, is the number of
kilopixels rasterized per second.}

\centering\begin{tabular}{>{\centering}p{0.25in}>{\centering}p{0.35in}>{\centering}p{0.25in}>{\centering}p{0.25in}>{\centering}p{0.25in}>{\centering}p{0.15in}>{\centering}p{0.25in}}
\toprule 
Pixels & Res\\
($\unit{Kpc/px}$) & $\bar{n}$ & CPUs & Wall-time\\
($\unit{hours}$) & $\eta$ & Rate\\
($\unit{K/sec}$)\tabularnewline
\midrule
5.6G & 58.4 & 80 & 128 & \multicolumn{1}{c}{1.63} & 1.45 & 11.3\tabularnewline
22.5G & 29.2 & 330 & 256 & 3.17 & 1.66 & 13.4\tabularnewline
90G & 14.6 & 1300 & 512 & 3.35 & 1.57 & 24.0\tabularnewline
90G & 14.6 & 1300 & 512 & 3.06 & 1.39 & 23.3\tabularnewline
360G & 7.3 & 5300 & 512 & 7.65 & 1.39 & 37.2\tabularnewline
1160G & 4.2 & 16000 & 1344 & 10.1 & 1.56 & 37.2\tabularnewline
\bottomrule
\end{tabular}
\end{table}

The rasterization of the images were carried out on
Blacklight at PSC. It is interesting to note that for the
largest images, the disk I/O wall-time, limited by the I/O
bandwidth of the machine, overwhelms the total computing
wall-time. The performance of the I/O subsystem shall be an
important factor in the selection of machines for data
visualization at this scale.

\section{Image and Animation Viewing}

Once large images or animation frames have been created,
viewing them presents a separate problem.  We use the
GigaPan technology for this, which enables someone with a
web browser and Internet connection to access the simulation
at high resolution. In this section we give a brief overview
of the use of the GigaPan viewer for exploring large static
images, as well as the recently developed GigaPan Time
Machine viewer for gigapixel animations.

\subsection{Gigapan}

Individual gigapixel-scale images are generally too large to
be shared in easy ways;  they are too large to attach to
emails, and may take minutes or longer to transfer in their
entirety over typical Internet connections.  GigaPan
addresses the problem of sharing and interactive viewing of
large single images by streaming in real-time the portions of
images actually needed by the viewer of the image, based on
the viewers current area of focus inside the image.  To
support this real-time streaming, the image is divided up and
rendered into small tiles of multiple resolutions.  The
viewer pulls in only the tiles needed for a given view.
Many mapping programs (e.g., Google Maps) use the same
technique.

\begin{figure*}[h!]
\centering
\includegraphics[scale=0.95]{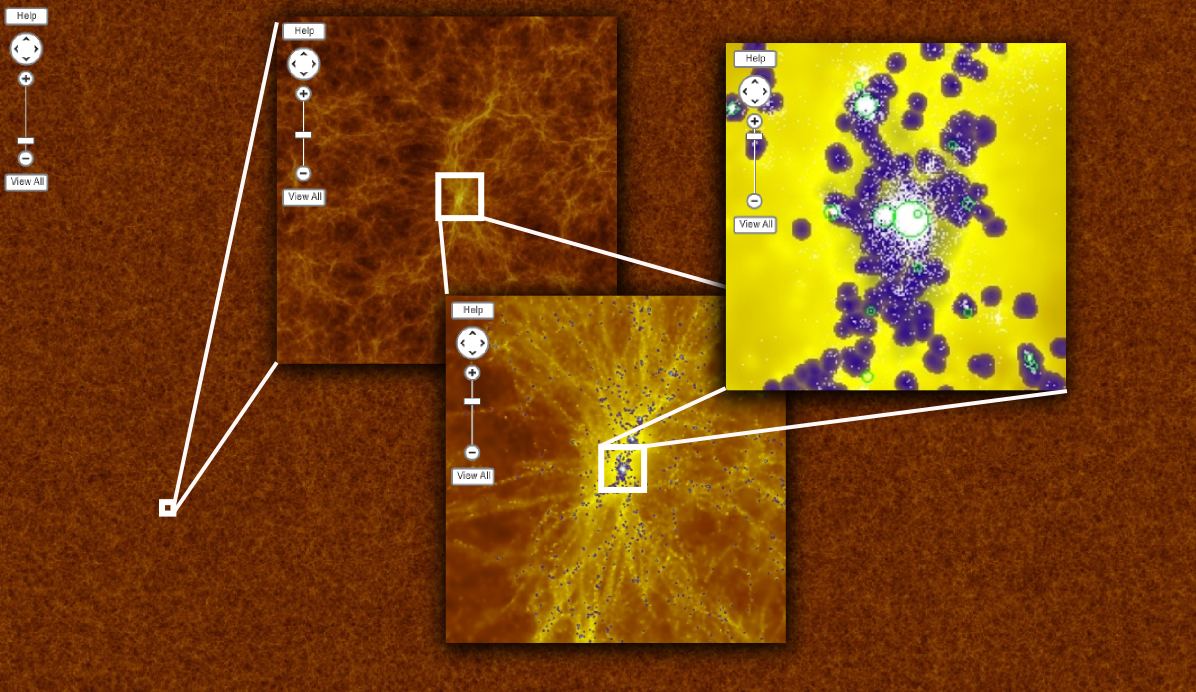}

\caption{GigaPan View of the MassiveBlack Simulation at
$z=4.75$\label{fig:GigapanView}.  The images are screen grab
from the GigaPan viewer: we have left the magnification bar
visible.  The background is an overall view of the entire
snapshot. We also show zooms into region around one of the
most massive blackholes in the simulation, as shown in the
right most zoom with the largest circle. The three zoom
levels are $\unit[80]{h^{-1}Mpc}$, $\unit[8]{h^{-1}Mpc}$,
$\unit[800]{h^{-1}Kpc}$, from left to right. }

\end{figure*}

\begin{figure*}[h!]
\centering
\includegraphics[scale=0.90]{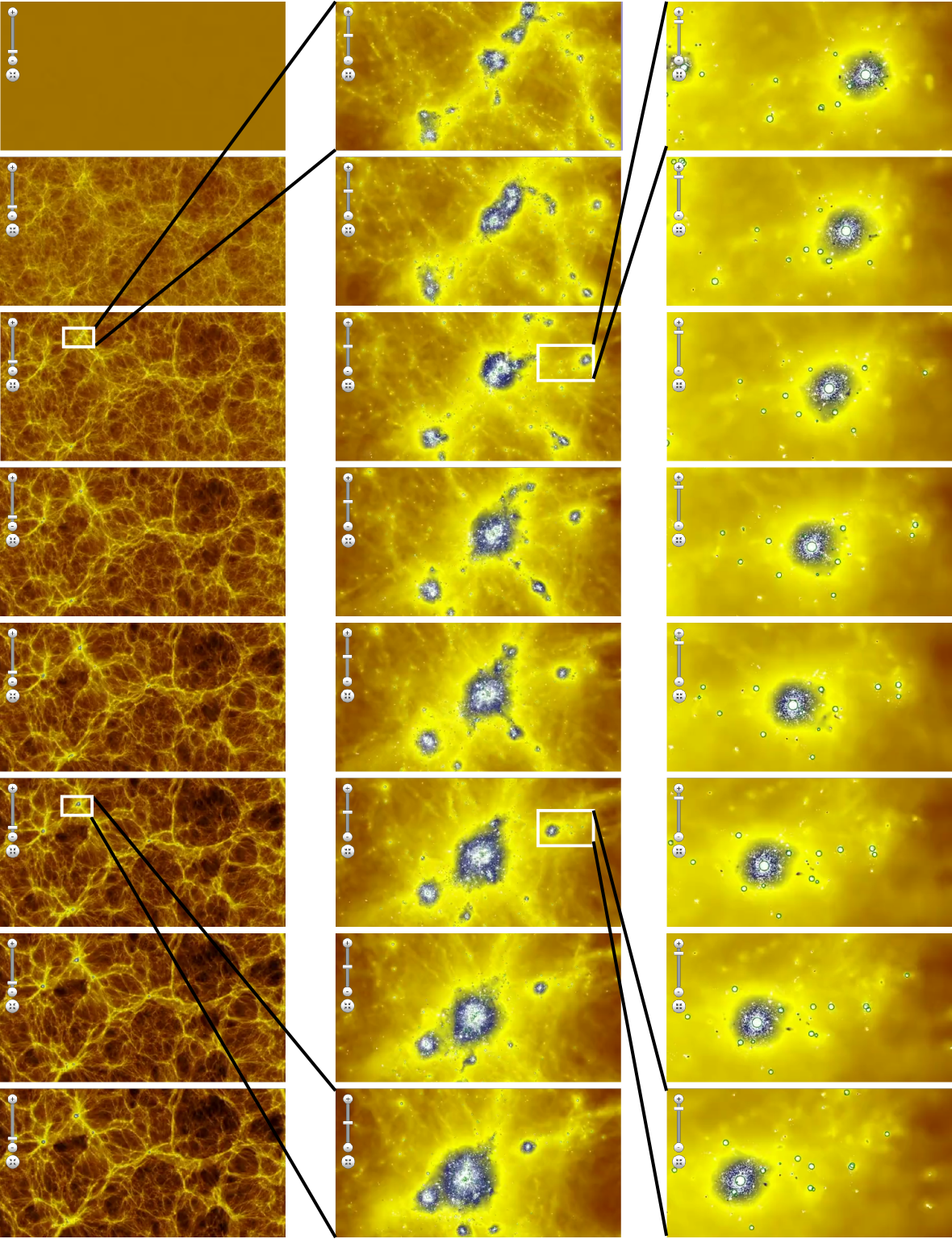}

\caption{GigaPan Time Machine Animation of the E5
simulation.  These are screen grabs from the GigaPan Time
Machine viewer. In the left column we show 8 frames (out of
1367 in the full animation) which illustrate the evolution
of the entire simulation volume (at time intervals of
$\unit[2]{Gyears}$.  The middle panel zooms in to show
formation of the largest halo through merger event. The
right panel shows some of the history of a smaller halo.}

\end{figure*}

We have uploaded an example terapixel image\footnote{image
at http://gigapan.org/gigapans/76215/} of the  redshift
$z=4.75$ snapshot of the MassiveBlack simulation to the
GigaPan website, which is run as a publicly accessible resource
for sharing and viewing large images and movies. The
dimension of the image is $1440000\times810000$, and the
finished image uncompressed occupies $\unit[3.58]{TB}$ of storage
space.  The compressed hierarchical data storage in Gigapan
format is about 15\% of the size, or $\unit[0.5]{TB}$.  There is no
fundamental limits to size, provided the data can be stored on the disk.
It is possible to create directly the compressed tiles of a GigaPan,
bypassing the uncompressed image as an intermediate step,
and thus reducing the requirement on memory and disk
storage. We leave this for  future work.

On the viewer side, static GigaPan works well at different
bandwidths; the interface remains responsive independent of
bandwidth, but the imagery resolves more slowly as the
bandwidth is reduced. \unit[250]{kilobits/sec} is a
recommended bandwidth for exploring with a $1024\times768$
window, but the system works well even when the bandwidth is
lower.

An illustration of the screen output is shown in
Figure \ref{fig:GigapanView}.  The reader is encouraged to
visit the website to explore the image.

\subsection{GigaPan Time Machine}

In order to make animations, one starts with the rendered
images of each individual snapshot in time. These can be
gigapixel in scale or more. In our example, using the E5
simulation (Section 2.2) we have 1367 images each with 0.75
gigapixels.

One approach to showing gigapixel imagery over time would be
to modify the single-image GigaPan viewer to animate by
switching between individual GigaPan tile-sets. However, this
approach is expensive in bandwidth and CPU, leading to
sluggish updates when moving through time.

To solve this problem, we created a gigapixel video
streaming and viewing system called GigaPan Time
Machine \citep{GIGAPAN}, which allows the user to fluidly
explore gigapixel-scale videos across both space and time.
We solve the bandwidth and CPU problems using an approach
similar to that used for individual GigaPan images:  we
divide the gigapixel-scale video spatially into many smaller
videos.  Different video tiles contain different locations
of the overall video stream, at different levels of detail.
Only the area currently being viewed on a client computer
need be streamed to the client and decoded.  As the user
translates and zooms through different areas in the video,
the viewer scales and translates the currently streaming
video, and over time the viewer requests from the server
different video tiles which more closely match the current
viewing area.  The viewing system is implemented in
Javascript+HTML, and takes advantage of recent browser's
ability to display and control videos through the new HTML5
video tag.  

The architecture of GigaPan Time Machine allows the content
of all video tiles to be precomputed on the server; clients
request these precomputed resources without additional CPU
load on the server. This allows scaling to many
simultaneously viewing clients, and allows standard caching
protocols in the browser and within the network itself to
improve the overall efficiency of the system. The minimum
bandwidth requirement to view videos without stalling
depends on the size of the viewer, the frame rate, and the
compression ratios. The individual videos in ``Evolution of
the Universe'' (the E5 simulation, see below for link)
are currently encoded at \unit[25]{FPS} with
relatively low compression. The large video tiles require a
continuous bandwidth of \unit[1.2]{megabits/sec}, and a
burst bandwidth of \unit[2.5]{megabits/sec}.

We have uploaded an example
animation\footnote{http://timemachine.gigapan.org/wiki/Evolution\_of\_the\_Universe}
of the E5 simulation, showing its evolution over the
interval between redshift $z=200$ and $z=0$ with 1367 frames
equally spaced in time by \unit[10]{Myr}. Again, the reader is
encouraged to visit the website to explore the image.

\section{Conclusions}

We have presented a framework for generating and viewing
large images and movies of the formation of structure in
cosmological SPH simulations. This framework has been
designed specifically to tackle the problems that occur with
the largest datasets. In the generation of images, it
includes parallel rasterization (for either shared and
distributed memory) and adaptive pixel filling which leads
to a well behaved filling rate at high resolution. For
viewing images, the GigaPan viewers use hierarchical caching
and cloud based storage to make even the largest of these
datasets fully explorable at high resolution by anyone with
an internet connection. We make our image making toolkit
publicly available, and the GigaPan web resources are
likewise publicly accessible.

\acknowledgements{}

This work was supported by NSF Awards OCI-0749212,
AST-1009781, and the Moore foundation. The following
computer resources were  used in this research:
\facility{Kraken (NICS)}, \facility{Blacklight (PSC)},
\facility{Warp (Carnegie Mellon University)}.  Development
of this work has made extensive use of the Bruce and Astrid
McWilliams eScience Video Facility at Carnegie Mellon
University.

\end{document}